\documentclass[aps,pra,showpacs,twocolumn]{revtex4-1}
\usepackage{bm}
\usepackage{amsmath}
\usepackage{slashed}
\usepackage{float}
\allowdisplaybreaks
\usepackage{subcaption}
\usepackage{dcolumn}
\usepackage{graphicx}
\newcolumntype{.}{D{x}{}{-1}}

\newcolumntype{w}[1]{D{.}{.}{#1}}
\newcolumntype{L}{>{$}l<{$}}

\usepackage{graphicx}
\usepackage{xcolor}
\usepackage{nicefrac}

\newcommand{\vare}{\varepsilon}

\newcommand{\pr}{^{\prime}}

\newcommand{\hx}{\hat{\bfx}}
\newcommand{\bfp}{{\bm p}}

\newcommand{\bfx}{{\bm x}}

\newcommand{\bfr}{{\bm r}}

\newcommand{\balpha}{{\mbox{\boldmath$\alpha$}}}
\newcommand{\bgamma}{{\mbox{\boldmath$\gamma$}}}
\newcommand{\bsigma}{{\mbox{\boldmath$\sigma$}}}

\newcommand{\intinf}{\int^{\infty}_{-\infty}}
\newcommand{\ka}{\kappa_a}
\newcommand{\kb}{\kappa_b}
\newcommand{\lbr}{\langle}
\newcommand{\rbr}{\rangle}
\newcommand{\ThreeJ}[6]{
        \left(
        \begin{array}{ccc}
        #1  & #2  & #3 \\
        #4  & #5  & #6 \\
        \end{array}
        \right)
        }
\newcommand{\SixJ}[6]{
        \left\{
        \begin{array}{ccc}
        #1  & #2  & #3 \\
        #4  & #5  & #6 \\
        \end{array}
        \right\}
        }
\newcommand{\NineJ}[9]{
        \left\{
        \begin{array}{ccc}
        #1  & #2  & #3 \\
        #4  & #5  & #6 \\
        #7  & #8  & #9 \\
        \end{array}
        \right\}
        }
\newcommand{\Dmatrix}[4]{
        \left(
        \begin{array}{cc}
        #1  & #2   \\
        #3  & #4   \\
        \end{array}
        \right)
        }

\newcommand{\Vcase}[2]{
        \left(
        \begin{array}{c}
        #1   \\
        #2   \\
        \end{array}
        \right)
        }

\newcommand{\Za}{{Z\alpha}}

\begin{document}

\title{Calculations of QED effects with the Dirac Green function}

\author{Vladimir A. Yerokhin}
\affiliation{Center for Advanced Studies, Peter the Great St.~Petersburg Polytechnic University,
Polytekhnicheskaya 29, 195251 St.~Petersburg, Russia}

\author{Anna V. Maiorova}
\affiliation{Center for Advanced Studies, Peter the Great St.~Petersburg Polytechnic University,
Polytekhnicheskaya 29, 195251 St.~Petersburg, Russia}

\begin{abstract}
Modern spectroscopic experiments in few-electron atoms reached the level of precision at which
an accurate description of quantum electrodynamics (QED) effects is mandatory. In many cases,
theoretical treatment of QED effects need to be performed without any expansion in the nuclear
binding strength parameter $Z\alpha$ (where $Z$ is the nuclear charge number and $\alpha$ is
the fine-structure constant). Such calculations involve multiple summations over the whole
spectrum of the Dirac equation in the presence of the binding nuclear field, which can be
evaluated in terms of the Dirac Green function. In this paper we describe the technique of
numerical calculations of QED corrections with the Dirac Green function, developed in numerous
investigations during the last two decades.
\end{abstract}

\maketitle

%
%
\section{Introduction}
\label{sec:intro}

Few-electron highly-charged ions are widely considered as important tools in testing quantum
electrodynamics (QED) theory in the presence of the binding nuclear field
\cite{mohr:98,beyer:03,indelicato:19}. Since the nuclear field in highly-charged ions is strong,
its binding strength cannot be used as an expansion parameter and theoretical investigations of
QED effects should be carried out to all orders in $\Za$, where $Z$ is the nuclear charge number
and $\alpha$ is the fine structure constant. This is achieved by working in the so-called Furry
picture, where the classical binding field of the nucleus is included into the zeroth-order
approximation.

Interaction of the electron(s) bound in the field of the nucleus with the quantized radiation
field gives rise to the QED effects, which are accounted for by an expansion in powers of
$\alpha$. General expressions for individual QED corrections are derived within the dedicated
methods, most notably, the adiabatic $S$-matrix formalism by Gell-Mann, Low and Sucher
\cite{gellmann:51,sucher:57} and by the two-time Green function method by Shabaev
\cite{shabaev:02:rep}.

The major difficulty encountered in calculations of QED corrections comes from the presence of
infinite summations over the whole spectrum of the Dirac equation with the binding nuclear
potential. These sums can be interpreted in terms of the so-called bound electron propagators, or
the Dirac Green function.

Calculations of QED effects with the Dirac Green functions started in 1970th with computations of
the one-loop self-energy~\cite{desiderio:71,mohr:74:a,mohr:74:b} and
vacuum-polarization~\cite{soff:88:vp,manakov:89:zhetp}. Over~the past years, the~number and the
complexity of QED calculations performed to all orders in the binding field has been increasing
rapidly. These calculations have been successful not only in improving the achievable precision
but also in extending the range of the studied effects, from~the classical Lamb shift to the QED
corrections to the hyperfine structure, the~$g$ factor, the~transition amplitudes, the~nuclear
magnetic shielding, etc. This progress was due to not merely the increased computing speed and
the availability of parallel computer resources, but~also due to the development of new
computational algorithms and~methods.

With the present work we summarize the computational technique developed for calculations of
various QED corrections with the bound-electron propagators, paying particular attention to the
notoriously problematic diagrams with several propagators inside the radiative photon loop. This
technique was developed in numerous calculations performed over the last two decades, notably, in
Refs.~\cite{yerokhin:99:pra,yerokhin:01:hfs,yerokhin:04,yerokhin:10:sehfs,yerokhin:18:sese}.

The relativistic units ($\hbar=c=m=1$) and the Heaviside charge units ($ \alpha = e^2/4\pi$,
$e<0$) will be used throughout this paper.

%
%
\section{Dirac Green function}
\label{sec:green}

The electron propagator $S(x_2,x_1)$ is standardly defined as the vacuum expectation value of the
time-ordered product of the electron-positron field operators,
\begin{equation} \label{pr1}
S(x_2,x_1) = -i \lbr 0| T\, \Psi(x_2)\, \overline{\Psi}(x_1)|0 \rbr \,,
\end{equation}
where $T$ denotes the time-ordered product, $\overline{\Psi} = \Psi^{\dag}\gamma^0$, and $\Psi$
is the electron-positron field operator (see, e.g., Ref.~\cite{mohr:98}),
\begin{equation} \label{pr2}
\Psi(x) = \sum_k \hat{a}_k \,\varphi_k^{(+)}+ \sum_k \hat{b}_k^{\dag}\, \varphi_k^{(-)}\ .
\end{equation}
Here, $\hat{a}^{\dag}$ ($\hat{b}^{\dag}$) and $\hat{a}$ ($\hat{b}$) are the electron (positron)
creation and annihilation operators, respectively; $\varphi_k^{(\pm)}(x) =
\psi^{(\pm)}_k(\bfx)\exp(-i\vare_k^{(\pm)}t)$ are single-particle electron (positron) states in
the external field $A(x)$, and $\psi_k^{(\pm)}$ are the positive- and negative-energy
eigenfunctions of the time-independent Dirac Hamiltonian ${\cal H}_D$,
\begin{equation}    \label{pr3}
{\cal H}_D\,\psi_k(\bfx) \equiv
 \bigl[ \balpha \cdot (\bfp-e{\bf A}) + \beta m+ eA^0 \bigr] \psi_k(\bfx) = \vare_k
\psi_k(\bfx) \,,
\end{equation}
where $\beta = \gamma^0$, $\balpha = \beta \bgamma$, and $x = (t,\bfx)$ is a four-vector.
Substituting Eq.~(\ref{pr2}) into Eq.~(\ref{pr1}), we get
\begin{align}    \label{pr4}
S(x_2,x_1) =&\  -i\theta(t_2-t_1)
\sum_k\varphi_k^{(+)}(x_2)\,\overline{\varphi_k^{(+)}}(x_1)
\nonumber \\ &
    +i\theta(t_1-t_2) \sum_k \varphi_k^{(-)}(x_2),\overline{\varphi_k^{(-)}}(x_1) \ ,
\end{align}
where $\theta(t)$ is the Heaviside step function. This expression can be conveniently rewritten
an equivalent form
\begin{equation}    \label{pr5}
S(x_2,x_1) = \frac{1}{2\pi} \intinf d\omega\, e^{-i\omega(t_2-t_1)} \sum_n
    \frac{\psi_n(\bfx_2)\,\overline{\psi}_n(\bfx_1)}{\omega-\vare_n(1-i0)} \,,
\end{equation}
where the summation is carried out over both positive and negative energy states. Equivalence of
these two representations for the electron propagator can be checked by performing the $\omega$
integration in Eq.~(\ref{pr5}) by Cauchy's theorem.

It can be easily shown (see, {\it e.g.}, Ref.~\cite{berestetskii:82:qed}) that the electron
propagator satisfies the differential equation
\begin{equation} \label{pr6}
\big[ i\slashed{\partial}_2 -e\slashed{A}(x_2)-m\big] S(x_2,x_1) = \delta^4(x_2-x_1)\,,
\end{equation}
where slashed symbols denote contractions with $\gamma$ matrices, $\slashed{\partial} =
\gamma_{\mu}\partial^{\mu}$ and $\slashed{A} = \gamma_{\mu}A^{\mu}$. In the absence of the
external field, this equation can be solved in a closed form. The result is the free-electron
propagator,
\begin{equation}    \label{pr7}
S^{(0)}(x_2-x_1) = \int \frac{d^4p}{(2\pi)^4}\, e^{-ip\cdot
    (x_2-x_1)}  \frac{\slashed{p}+m}{p^2-m^2+i0} \,,
\end{equation}
where $p = (p^0,\bfp)$ is a four-vector.

Within the Feynman-diagram technique (see, e.g., Ref.~\cite{shabaev:02:rep}), the integration
over the time components of the arguments of the electron propagator is usually carried out in
the general form, so that in practical calculations one deals with the Fourier transform of
$S(x_2,x_1)$ with respect to the time variable $\tau = t_2-t_1$. The result is referred to as the
Dirac Green function,
\begin{align}    \label{pr12}
G(E,\bfx_2,\bfx_1) &\ =  \intinf d\tau\, e^{iE\tau}\, S(x_2,x_1)\gamma^0
 \nonumber \\ &
= \sum_n
        \frac{\psi_n(\bfx_2)\,\psi^{\dag}_n(\bfx_1)}{E-\vare_n(1-i0)} \,.
\end{align}
From Eq.~(\ref{pr6}), we deduce that $G(E,\bfx_2,\bfx_1)$ satisfies the differential equation
\begin{equation}    \label{pr13}
(E-{\cal H}_D)\, G(E,\bfx_2,\bfx_1) = \delta^3(\bfx_2-\bfx_1) \,,
\end{equation}
where ${\cal H}_D$ is the Dirac Hamiltonian, see Eq.~(\ref{pr3}).

In this work we are interested in the Dirac Green function in a central field. In this case the
angular structure of $G(E,\bfx_2,\bfx_1)$ follows from Eq.~(\ref{pr12}) and from the angular
dependence of the Dirac solutions \cite{rose:61},
\begin{equation}    \label{pr14}
\psi_{n}(\bfx)
        =\left(  {g_n(x)\, \chi_{ \kappa_n \mu_n}(\hat{\bfx})}
           \atop{if_n(x)\, \chi_{-\kappa_n \mu_n}}(\hat{\bfx}) \right) \,,
\end{equation}
where $g_n(r)$ and $f_n(r)$ are the upper and the lower radial components of the wave function,
respectively, $\chi_{\kappa \mu}(\hat{\bfx})$ is the spin-angular spinor, $\kappa$ is the
relativistic angular-momentum quantum number, $\mu$ is the angular momentum projection, $x =
|\bfx|$, and $\hat{\bfx} = \bfx/|\bfx|$. We thus obtain the standard partial-wave representation
of the Dirac Green function \cite{wichmann:56,brown:56}
\begin{widetext}
\begin{equation}    \label{pr15}
G(E,\bfx_2,\bfx_1) = \sum_{\kappa}
    \Dmatrix{G^{11}_{\kappa}(E,x_2,x_1)\,\pi_{\kappa}^{++}(\hx_1,\hx_2)}
            {-i\,G^{12}_{\kappa}(E,x_2,x_1)\,\pi_{\kappa}^{+-}(\hx_1,\hx_2)}
            { i\,G^{21}_{\kappa}(E,x_2,x_1)\,\pi_{\kappa}^{-+}(\hx_1,\hx_2)}
            {  G^{22}_{\kappa}(E,x_2,x_1)\,\pi_{\kappa}^{--}(\hx_1,\hx_2)} \,,
\end{equation}
where $\pi_{\kappa}^{\pm\pm}(\hx_1,\hx_2) = \sum_{\mu}\chi_{\pm\kappa\mu}(\hx_1)
\,\chi_{\pm\kappa\mu}^{\dag}(\hx_2)$, and $G^{ij}_{\kappa}(E,x_2,x_1)$ are the radial components
of the Dirac Green function.

For a static potential [$eA^0(\bfx) = V(x)$, ${\bf A}(x) = 0$], Eq.~(\ref{pr13}) in the matrix
form reads
\begin{equation}    \label{g1}
\Dmatrix{E-m-V(x)} {-(\bsigma \cdot\bfp)}{-(\bsigma \cdot \bfp)}
        {E+m-V(x)}
     G(E,\bfx,\bfx\pr) = \delta(\bfx-\bfx\pr)\, I \,,
\end{equation}
where $I$ is the 2$\times$2 identity matrix. Substituting Eq.~(\ref{pr15}) and using the
identities
\begin{equation}   \label{g2}
(\bsigma \cdot \bfp)\, f(x)\, \chi_{\kappa \mu}(\hx) = i \Bigr(
    \frac{\partial}{\partial x}+ \frac{1+\kappa}{x} \Bigl)\, f(x)\, \chi_{-\kappa
    \mu}(\hx) \,,
\end{equation}
and
\begin{equation}    \label{g2a}
 \delta(\bfx-\bfx\pr) = \frac1{xx\pr}\, \delta(x-x\pr)\, \sum_{\kappa \mu}
    \chi_{\kappa \mu}(\hx) \chi_{\kappa \mu}^{\dag}(\hx\pr) \,,
\end{equation}
we obtain the equation for the radial Dirac Green function,
\begin{equation}    \label{g3}
\big( E\,I - h_{D,\kappa}\big)\,G_{\kappa}(E,x,x\pr)  \equiv
\Dmatrix{\displaystyle E-m-V(x)} {\displaystyle \frac{ d}{ d
x}-\frac{\kappa-1}{ x}} {\displaystyle -\frac{
d}{ d x}-
    \frac{\kappa+1}{x}}
    {\displaystyle E+m-V(x)   }
     G_{\kappa}(E,x,x\pr) = \frac{1}{xx\pr}\,\delta(x-x\pr)\,I \,,
\end{equation}
where $h_{D,\kappa}$ is the radial Dirac Hamiltonian and $G_{\kappa}$ is the 2$\times$2 matrix of
radial components of the Green function $G^{ij}_{\kappa}$, defined by Eq.~(\ref{pr15}).
\end{widetext}

\subsection{Representation in terms of regular and irregular solutions}
\label{sec:WhittGreen}

The solution of an inhomogeneous differential equation (\ref{g3}) can be constructed from the
solutions of the corresponding homogeneous equation bounded at infinity
($\phi_{\kappa}^{\infty}$) and at origin ($\phi_{\kappa}^0$),
\begin{align} \label{g4}
G_{\kappa}(E, x,x\pr) = &\
\frac1{\Delta_{\kappa}(E)}\,\bigg[
\phi^{\infty}_{\kappa}(x)\, \phi^{0^{T}}_{\kappa}(x\pr)\, \theta(x-x\pr)
 \nonumber \\ &
+     \phi^{0}_{\kappa}(x)\, \phi^{\infty^{T}}_{\kappa}(x\pr)\, \theta(x\pr-x)\bigg]
    \,,
\end{align}
where the subscript $T$ denotes the transposition, $\phi^0_{\kappa}$ and $\phi^{\infty}_{\kappa}$
are the two-component solutions of the homogeneous radial Dirac equation, and
$\Delta_{\kappa}(E)$ is their Wronskian,
\begin{align}    \label{g5}
\Delta_{\kappa}(E) = x^2 \phi^{0^{T}}_{\kappa}(x)
\Dmatrix{\ \ 0}{1}{-1}{0}
\phi^{\infty}_{\kappa}(x)
\,,
\end{align}
which is independent on $x$. When the energy parameter $E$ of the Green function is an eigenvalue
of the Dirac Hamiltonian, the two solutions $\phi^0_{\kappa}$ and $\phi^{\infty}_{\kappa}$
coincide (up to a constant factor) and their Wronskian vanishes, $\Delta_{\kappa}(E_n) = 0$. This
gives rise to poles of the Green function. The Green function has also branch points at $E=\pm
m$, with cuts along the real axis for $|E|>m$, as will be discussed in more details below.

For the point-nucleus Coulomb potential [$V(x) = -Z\alpha/x$] the equation (\ref{g3}) can be
solved analytically \cite{wichmann:56} in terms of the Whittaker functions. The result is
commonly referred to as the Dirac-Coulomb Green function. The radial Dirac-Coulomb Green function
is represented by the form (\ref{g4}), with the functions $\phi^0$ and $\phi^{\infty}$ given by
\cite{mohr:74:a}
\begin{equation}
\label{gc1}
\phi^0_C(x) = \left( {\phi^{0}_{C,+}(x)}\atop{\phi^{0}_{C,-}(x)} \right) \,, \ \ \
\phi^{\infty}_C(x) = \left( {\phi^{\infty}_{C,+}(x)}\atop{\phi^{\infty}_{C,-}(x)} \right) \ ,
\end{equation}
\begin{eqnarray}  \label{gc2} \phi_{C,\pm}^{0}(x) &=&
        \frac{\sqrt{1\pm \varepsilon}}{x^{3/2}}
        \bigg[ (\lambda-\nu)M_{\nu-(\nicefrac12),\,\lambda}(2cx)
        \nonumber \\ &&
        \mp \left(\kappa-\frac{\alpha Z}{c}\right) M_{\nu+(\nicefrac12),\,\lambda}(2cx) \bigg]
    \,, \\
\phi_{C,\pm}^{\infty}(x) &=&          \label{gc2a}
        \frac{\sqrt{1\pm \varepsilon}}{x^{3/2}}
        \bigg[ \left(\kappa+\frac{\alpha Z}{c}\right)W_{\nu-(\nicefrac12),\,\lambda}(2cx)
        \nonumber \\ &&
        \pm W_{\nu+(\nicefrac12),\,\lambda}(2cx) \bigg] \,,
\end{eqnarray}
and
\begin{equation}    \label{gc3}
\Delta_{C,\kappa}( E) = -4c^2 \frac{\Gamma(1+2\lambda)}{\Gamma(\lambda-\nu)} \,,
\end{equation}
where $\varepsilon = E/m$, $c=\sqrt{1- \varepsilon^2}$, $\lambda=\sqrt{\kappa^2-(\alpha Z)^2}$,
$\nu=Z\alpha\, \varepsilon/c$, and $M_{\alpha,\beta}$ and $W_{\alpha,\beta}$ are the Whittaker
functions of the first and the second kind \cite{gradshteyn}, respectively. We mention the
opposite sign of the present definition of the Green function as compared to the definition of
Refs.~\cite{mohr:74:a,mohr:98}.

Zeros of the Wronskian (\ref{gc3}) correspond to the bound-state energy levels, $\lambda-\nu =
-n_r$ ($n_r = 0,1,\ldots$ is the radial quantum number), which yields the well-known formula for
the Dirac bound energies,
\begin{equation}    \label{gc4}
E_{\kappa,n_r} = m \left[
1+ \left(\frac{\alpha
                    Z}{\lambda+n_r}\right)^2\right]^{-1/2}\,.
\end{equation}

The cut structure of the Dirac-Coulomb Green function is defined by that of the square root
$\sqrt{m^2-E^2}$. The square root function is defined to be positive in the gap $-m < E < m$ on
the real $E$-axis. Outside of the gap, the sign of the square root is fixed by the condition
${\rm Re} (\sqrt{m^2-E^2}) > 0$. Special care should be taken evaluating the Green function for
real energies $|E|>m$. Behaviour of the Green function on the real $E$ axis is defined by the
sign of the infinitesimal addition in the energy denominator of Eqs. (\ref{pr5}) and
(\ref{pr12}). In our case the addition is negative and, therefore, the cut $E>m$ should be
approached from the upper half of the $E$ plane, and the cut $E<-m$ from the lower half. So,
e.g., starting from the gap $-m < E < m$ and approaching the branch cut $E> m$ from the upper
half-plane, we have the following prescription \cite{shabaev:00:rec} for the analytical
continuation of the square root: $\sqrt{m^2-E^2} \to -i \, \sqrt{E^2-m^2}$.

In the limit of $Z \to 0$, the Dirac-Coulomb Green function is reduced to the free Dirac Green
function. The corresponding radial solutions are given by \cite{mohr:74:a}
\begin{eqnarray}  \label{gf2}
\phi_{F,\pm}^{0}(x) &=&
 \Vcase{1}{i\frac{\kappa}{|\kappa|}}  \sqrt{1 \pm \varepsilon}\,j_{l_{\pm\kappa}}(icx)
    \,, \\
\phi_{F,\pm}^{\infty}(x) &=&          \label{gf2a}
         \Vcase{1}{i\frac{\kappa}{|\kappa|}}  \sqrt{1 \pm \varepsilon}\,h^{(1)}_{l_{\pm\kappa}}(icx)
    \,,
\end{eqnarray}
where $l_{\kappa} = |\kappa+1/2|-1/2$, $j(z)$ and $h^{(1)}(z)$ are spherical Bessel functions,
and in $\left( \cdots\right)$ the upper value is chosen for the ``+'' component and the lower,
for the ``-'' component. The Wronskian of the above solutions is $\Delta_{F,\kappa}(E) = 1/c$.

The numerical computation of the Whittaker functions required in calculations of QED corrections
with the Dirac-Coulomb Green function was first tackled by Mohr in
Refs.~\cite{mohr:74:a,mohr:74:b} (see also the review \cite{mohr:98}). His approach enabled an
accurate computation of the Whittaker functions in a wide range of arguments, including high
values of the relativistic angular parameter $\kappa$. A disadvantage of this numerical approach
was that it required the extended-precision arithmetic to be used in a certain range of the
arguments. A more economical variation of this approach was reported in
Ref.~\cite{yerokhin:99:pra}. It allowed a computation of Whittaker functions within the standard
double-precision arithmetics, for not very high partial waves ($|\kappa|  \lesssim 40$), which
turned out to be sufficient for many practical applications.

The representation (\ref{g4}) can also be used in computations of the Dirac Green function for
potentials other than the point-Coulomb potential. In~particular, ref.~\cite{soff:88:vp}
presented a numerical approach for computing the Dirac Green function for the potential induced
by the nuclear charge distribution given by the shell nuclear model $\rho(r) \propto
\delta(r-R)$. The~computation of the Dirac Green function for the homogeneously charged nuclear
model $\rho(r) \propto \theta(r-R)$ was reported in ref.~\cite{mohr:98}.

In practical calculations, more realistic models of the nuclear-charge distribution are often
required, first of all, the~two-parameter Fermi distribution. A~numerical approach for the
computation of the Dirac Green function with the spherically-symmetric Fermi nuclear model was
described in ref.~\cite{yerokhin:11:fns}. This approach can be easily generated for the case of
an arbitrary central potential approaching the Coulomb potential in the limit of $r\to \infty$,
in particular, for~a wide class of screened nuclear~potentials.

\begin{figure}[t]
\centerline{\resizebox{0.4\textwidth}{!}{\includegraphics{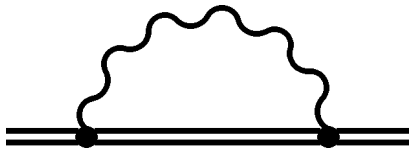}}}
 \caption{The one-loop self-energy correction. The double line represents the electron
 propagating in the binding field of the nucleus. The wavy line denotes the virtual photon.
 \label{fig:se}}
\end{figure}
\begin{figure}[t]
\centerline{\resizebox{0.25\textwidth}{!}{\includegraphics{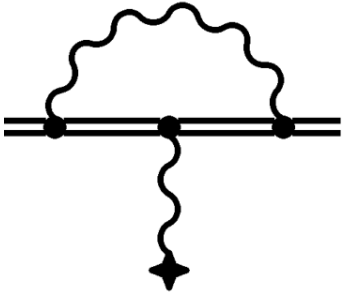}}}
 \caption{The magnetic-vertex self-energy correction. The wavy line terminated
 by a cross denotes the interaction with an external magnetic field.
 \label{fig:vermag}}
\end{figure}
\begin{figure}[t]
\centerline{\resizebox{0.3\textwidth}{!}{\includegraphics{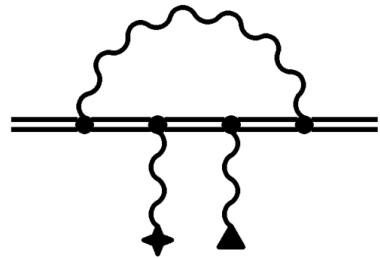}}}
 \caption{The double-vertex self-energy correction. The wavy line terminated
 by a triangle denotes the hyperfine interaction.
 \label{fig:dblver}}
\end{figure}

\subsection{Finite basis set representations}
\label{sec:Bsplines}

Using the spectral representation of the Green function (\ref{pr12}), we can represent the radial
Dirac Green function as
\begin{equation}    \label{g6}
G_{\kappa}(E, x,x\pr) = \sum_n \frac{\phi_{\kappa,n}(x)\,\phi_{\kappa,n}^T(x\pr)}
        {E-\vare_{\kappa,n}} \,,
\end{equation}
where $\phi_{\kappa, n}$ are the two-component radial Dirac functions with energies
$\vare_{\kappa,n}$ satisfying the radial Dirac equation
\begin{equation}    \label{g7}
h_{D,\kappa}\,\phi_{\kappa, n}(x) = \vare_{\kappa,n}\,\phi_{\kappa, n}(x) \,.
\end{equation}
The sum over $n$  in Eq.~(\ref{g6}) should be understood as a summation over the discrete part of
the spectrum and the integration over the positive and negative continuum parts of the spectrum.

A very useful approach to the numerical evaluation of the Dirac Green function is provided by the
finite basis set method. In this method, the radial Dirac solutions are approximately represented
by linear combinations of (a finite set of) two-component basis functions $u_i(x)$\,,
\begin{align}
\phi_{n}(x) =  \sum_{i=1}^{N} c_i\,u_i(x)\,.
\end{align}
Within this representation, the solution of the radial Dirac equation (\ref{g7}) is reduced to a
generalized eigenvalue problem for the coefficients $c_i$,
\begin{align} \label{eq24a}
\frac12 \big[ \lbr u_i|h_{D}|u_k\rbr + \lbr u_k|h_{D}|u_i\rbr\big] c_k = E\,\lbr u_i|u_k\rbr\,c_k\,,
\end{align}
where the summation over repeated indices is implied and $i,k = 1\ldots N$. This equation can be
solved numerically by the standard methods of linear algebra, which yields the set of $N$
eigenvectors and eigenvalues of the radial Dirac equation. After that, by using Eq.~(\ref{g6})
one obtains a finite basis-set representation of the radial Dirac Green function.

The choice of the basis function $u_k$ can vary. One of the most successful implementations is
delivered by the dual kinetic balance (DKB) basis \cite{shabaev:04:DKB} constructed with
$B$-splines \cite{deboor:78}. Within this method, the radial Dirac solutions are represented as
\begin{align} \label{eq24}
\phi_{\kappa,n}(x)
 = &\ \sum_{i = 1}^{\nicefrac{N}{2}} c_i\,
\Vcase{B_i(x)}{\displaystyle \frac1{2m}\biggl(\frac{d}{dx}+\frac{\kappa}{x}\biggl)\,B_i(x)}
 \nonumber \\ &
+ \sum_{i = 1}^{\nicefrac{N}{2}} c_{i+\nicefrac{N}{2}}\, \Vcase{\displaystyle
\frac1{2m}\biggl(\frac{d}{dx}-\frac{\kappa}{x}\biggl)\,B_i(x)}{B_i(x)}\,,
\end{align}
where $\left\{B_i(x)\right\}_{i = 1}^{\nicefrac{N}{2}}$ is the set of $B$-splines
\cite{deboor:78} on the interval $(0, R)$, where $R$ is the cavity radius, chosen to be
sufficiently large in order to have no influence on the calculated properties of the atom. The
$B$-splines are chosen to vanish at $x = 0$ and $x = R$, thus yielding the zero boundary
conditions for the wave functions, $\phi(0) = \phi(R) = 0$.

It needs to be stressed that the DKB anzatz (\ref{eq24}) assumes that the potential in the Dirac
equation is regular at $r\to 0$. This means that it can be used for solving the Dirac equation
for an extended-nucleus potential, but {\it not} for the point-nucleus Coulomb potential. The
advantages of the DKB basis is the absence of the so-called spurious states, the correct
behaviour of the upper and lower radial components at $r\to 0$ and, as a consequence, an improved
convergence of the calculated atomic properties with increase of the size of the basis set.

For the point nuclear model, one often uses the simpler anzatz of Ref.~\cite{johnson:88},
\begin{align} \label{eq25}
\phi_{\kappa,n}(x)
 = &\ \sum_{i = 1}^{\nicefrac{N}{2}} c_i\,
\Vcase{B_i(x)}{0}
 \nonumber \\ &
+ \sum_{i = 1}^{\nicefrac{N}{2}} c_{i+\nicefrac{N}{2}}\,
\Vcase{0}{B_i(x)}\,.
\end{align}
It should be noted that the anzatz (\ref{eq25}) leads to appearance of spurious eigenstates
(highly oscillating eigenvectors with unphysical energies), as analytically proved in
Ref.~\cite{shabaev:04:DKB}. In practical calculations, these spurious states do not cause
significant problems (since their contributions to integrals is very small due to rapid
oscillations), but their presence is manifested in a slower convergence of the calculated results
as $N\to \infty$.

We also mention the space-discretization method for the solution of the Dirac equation
\cite{salomonson:89:pair1,salomonson:89:pair2}, which can be regarded as a variant of the finite
basis-set method with the basis constructed with $\delta$-functions. For practical calculations,
the $B$-spline basis has the advantages of being more compact and consisting of continuous
functions, while eigenvectors in the space-discretization method are defined on a grid only.
However, the space-discretization method was successfully used in many calculations of QED
effects by G\"oteburg group, notably, in Refs.~\cite{lindgren:95:pra,persson:97:g,asen:02}.
Moreover, this method apparently yields a better convergence than the $B$-spline approach in
calculations of the Wichmann-Kroll vacuum-polarization corrections \cite{persson:93:vp}.

\subsection{Discussion }

In Sec.~\ref{sec:WhittGreen} and \ref{sec:Bsplines} we described the two main representations of
the bound electron propagator used in modern calculations of QED effects in atomic spectra. The
first one is the representation in terms of the regular and irregular Dirac solutions (in what
follows, the Green's function approach) and the second is the finite basis set method. The other
known representations of the Dirac-Coulomb Green function are not discussed in the present work,
since they have not been proved useful in the calculations we consider here. In particular, the
Sturmian expansion of the Dirac-Green function, widely used in the literature for the description
of multiphoton processes (see, {\em e.g.},
Refs.~\cite{zon:72,manakov:73,szymanowski:97,szmytkowski:97}), does not seem to be useful for the
calculations considered here. The main reasons are the numerical character of calculations and
the lack of convergence of the Sturmian expansion when the energy argument is in the complex
plane.

We now give a comparative discussion of the two main approaches. The basis-set method has several
attractive features. The corresponding numerical routine is relatively simple, flexible and can
easily incorporate any spherical-symmetric potential. Moreover, this method allows one to perform
summations over a part of the Dirac spectrum (e.g., over the positive or the negative part only)
and evaluate sums over spectrum with energy denominators different from the one in
Eq.~(\ref{g6}).

Another attractive feature of the basis-set method is that it provides an approximation to the
Green function which is a {\em continuous} function of the radial arguments at $x \approx x\pr$.
This is not the case for the exact Green function (\ref{g4}), whose components contain the
discontinuous step function $\theta(x-x\pr)$ (which yields a $\delta$-function in Eq.~(\ref{g3})
after differentiation). This feature is often referred to as the {\it radial ordering}, since the
exact Green function depends on $x_{<}$ and $x_{>}$, rather than just on $x$ and $x\pr$. This
feature complicates the numerical evaluation of matrix elements, especially for higher-order
diagrams with multiple radial integrations.

The basis-set method has also some important draw-backs as compared to the Green-function
approach. It has an additional parameter, the number of basis functions $N$, and the final result
should be investigated for stability when $N$ is increased. In practice, the dependence on the
basis size often sets a limitation on the accuracy of calculations. In addition, the number of
partial waves (i.e., the maximal value of $|\kappa|$) included in the numerical evaluation is
rather limited in the basis-set method. The typical number of partial waves employed in actual
calculations with the basis-set method is $\sim$20, while in the Green-function approach it can
be of order $10^4$ and more.

We conclude that the basis-set method has computational advantages for a restricted (but
sufficiently broad) class of problems, where the partial-wave expansion is well converging and
(or) the required numerical accuracy is not very high. The Green-function approach is preferable
for problems where (i) high numerical accuracy is needed, (ii) large numerical cancellations
occur, (iii) the partial-wave expansion does not converge rapidly, (iv) the contribution of
high-energy intermediate electron states is enhanced, leading to slow convergence of the
basis-set calculations with respect to $N$.

We now mention some of the calculations of QED corrections which used the above-mentioned methods
for computing the bound electron propagators. Historically, the~first was the Green-function
approach elaborated, most notably, by~Mohr in refs.~\cite{mohr:74:a,mohr:74:b}. This method was
developed further in calculations of the one-loop
self-energy~\cite{indelicato:92:se,jentschura:99:prl,yerokhin:99:pra,jentschura:01:pra,bigot:01:se},
the~ self-energy correction to the hyperfine splitting and $g$
factor~\cite{yerokhin:01:hfs,yerokhin:02:prl,yerokhin:08:prl}, the~screened QED
corrections~\cite{yerokhin:95,artemyev:97}, and~the QED corrections to the magnetic
shielding~\cite{yerokhin:11:prl}. The~$B$-spline basis-set method was used in calculations of the
two-photon exchange diagrams~\cite{blundell:93:b,yerokhin:00:prl,mohr:00:pra,volotka:14}, the~
one-loop self-energy~\cite{blundell:91:se}, the~nuclear
recoil~\cite{artemyev:95:pra,yerokhin:15:recprl,shabaev:17:prl}, the~screened QED
corrections~\cite{volotka:09,glazov:10}. The~space-discretization method was extensively applied
by the G\"oteburg group in calculations of the first-order self-energy and vacuum
polarization~\cite{persson:93:vp}, two-photon exchange~\cite{lindgren:95:pra,asen:02},
the~self-energy corrections to the bound-electron $g$ factor~\cite{persson:97:g}, to~the
hyperfine structure~\cite{sunnergren:98:pra}, to~the electron-electron
interaction~\cite{persson:96:2el}.

\section{General formulas}

In the present work we will consider actual calculations of three contributions originating from
the electron self-energy, specifically, the~matrix elements of the self-energy operator, the~
magnetic vertex operator, and~the double vertex operator, graphically represented in
Figures~\ref{fig:se}--\ref{fig:dblver}. The~corresponding diagrams involve one, two, and~three
bound electron propagators in the radiative photon loop, respectively. Calculations of
self-energy diagrams with one electron propagator started already in
1970th~\cite{desiderio:71,mohr:74:a,mohr:74:b}. First calculations of the vertex diagrams with
two electron propagators were performed in
1990th~\cite{indelicato:91:tca,yerokhin:95,persson:96:hfs,blundell:97:prl,yerokhin:96:pisma},
whereas the double vertex diagrams have been tackled only relatively
recently~\mbox{\cite{shabaev:05:prl,yerokhin:11:prl}}. There have been no calculations of
diagrams with more than three bound electron propagators in the radiative loop performed so~far.

The matrix element of the one-loop self-energy operator depicted on Fig.~\ref{fig:se} yields the
dominant contribution to the Lamb shift of the energy levels. It is given by
\begin{align}\label{gen:1}
\lbr a | \Sigma(\vare_a)| a\rbr = \frac{i}{2\pi}\intinf d\omega \sum_n
  \frac{\lbr an| I(\omega)| na\rbr}{\vare_a-\omega - u\vare_n}\,,
\end{align}
where the summation over $n$ is extended over the complete spectrum of the Dirac equation and $u
\equiv 1-i0$ ensures the positions of the singularities of the Green function with respect to the
integration contour. $I(\omega)$ is the operator of the electron-electron interaction, defined as
\begin{equation}\label{gen:2}
  I(\omega,\bfr_{1},\bfr_{2}) = e^2\, \alpha_{1}^{\mu} \alpha_{2}^{\nu}\, D_{\mu\nu}(\omega,\bfr_{12})\,,
\end{equation}
where $\alpha^{\mu} = (1,\balpha)$ are the Dirac matrices, $\bfr_{12} = \bfr_{1} - \bfr_{2}$, and
$D_{\mu\nu}(\omega,\bfr_{12})$ is the photon propagator. The photon propagator takes the simplest
form in the Feynman gauge, where it is given by
\begin{equation}\label{gen:3}
  D_{\mu\nu}(\omega,\bfr_{12}) = g_{\mu\nu}\,
  \frac{e^{i\sqrt{\omega^2+i0}\,r_{12}}}{4\pi r_{12}}\,,
\end{equation}
with $r_{12} = |\bfr_{12}|$.

\begin{widetext}
The matrix element of the magnetic vertex operator depicted in Fig.~\ref{fig:vermag} is the most
problematic part of the self-energy correction to the $g$ factor \cite{yerokhin:04}. The magnetic
vertex operator, accompanied by the corresponding reducible part, is defined by its matrix
elements as
\begin{align}\label{gen:4}
\lbr a |  \Lambda_{\rm vr}(\vare_a) | a\rbr
= &\
\frac{i}{2\pi}
  \intinf d\omega \,
\sum_{n_1n_2}
\frac{\lbr an_2| I(\omega)|n_1a\rbr\,\big[\lbr n_1|V_{g}|n_2\rbr - \lbr n_1|n_2\rbr\, \lbr a|V_g|a \rbr\big] }
  {(\vare_a-\omega-u\,\vare_{n_1})(\vare_a-\omega-u\,\vare_{n_2})}
\,,
\end{align}
where $V_g$ is the effective magnetic operator responsible for the $g$ factor \cite{yerokhin:04},
$V_g = (1/\mu_a)\,[\bm{r}\times\balpha ]_z$, with $\mu_a$ being the angular momentum projection
of the reference state $a$. We note that the scalar product $\lbr n_1|n_2\rbr$ in
Eq.~(\ref{gen:4}) can be trivially performed due to the orthogonality of the wave functions,
$\lbr n_1|n_2\rbr = \delta_{n_1n_2}$, but we find it convenient to keep it in the integral form.

The double-vertex operator matrix element shown in Fig.~\ref{fig:dblver} is the most problematic
part of the self-energy correction to the nuclear shielding
\cite{yerokhin:11:prl,yerokhin:12:shield}. It is defined, together with the corresponding
reducible parts, as
\begin{align}\label{gen:5}
\big< a \big| \Lambda_{\rm dvr}(\vare_a)\big| a\big>
=
2\,\frac{i}{2\pi}
  \intinf d\omega \,&\, \Bigg\{
\sum_{n_1n_2n_3}
\frac{\lbr an_3| I(\omega)|n_1a\rbr}
  {(\vare_a-\omega-u\,\vare_{n_1})(\vare_a-\omega-u\,\vare_{n_2})(\vare_a-\omega-u\,\vare_{n_3})}
 \nonumber \\
 & \times
 \Big[
   \lbr n_1|V_{g}|n_2\rbr\,\lbr n_2|V_{\rm hfs}|n_3\rbr
  -\lbr n_1|n_2\rbr\,\lbr n_2|V_{\rm hfs}|n_3\rbr\, \lbr a|V_{g}|a\rbr
 \nonumber \\
 &
 -\lbr n_1|V_{g}|n_2\rbr\,\lbr n_2|n_3\rbr \,\lbr a|V_{\rm hfs}|a\rbr
 +\lbr n_1|n_2\rbr\,\lbr n_2|n_3\rbr\, \lbr a|V_{g}|a\rbr\,\lbr a|V_{\rm hfs}|a\rbr
 \Big]
 \nonumber \\
 & -
\sum_{\mu_{a'} n_2}
 \frac{\lbr aa'| I(\omega)|a'a\rbr}{(-\omega + i0)^2}\,
 \lbr a|V_{g}|n_2\rbr\,\frac1{\vare_a-\vare_{n_2}}\lbr n_2|V_{\rm hfs}|a\rbr
 \Bigg\}
 \,,
\end{align}
where $V_{\rm hfs}$ is the effective magnetic operator responsible for the hyperfine interaction
\cite{yerokhin:10:sehfs}, $V_{\rm hfs} = (1/\mu_a)\,[\bm{r}\times\balpha ]_z/r^3$, $a'$ denotes
the reference state $a$ with a different angular momentum projection ($\mu_{a'}$), and the factor
of 2 in the front accounts for two equivalent diagrams.
\end{widetext}

The above general formulas for the self-energy and magnetic-vertex matrix elements contain
ultraviolet (UV) divergences. The standard approach to handle them  \cite{snyderman:91} is to
separate out one or two first terms of the expansion of the electron propagators in terms of the
interaction with the binding nuclear field. In order to get UV-finite results, the self-energy
operator needs a subtraction of the two first terms of the potential expansion,
\begin{align}\label{gen:6}
\Sigma (\vare_a) \to  \Sigma^{(2+)} (\vare_a) =
  \Sigma (\vare_a) - \Sigma^{(0)} (\vare_a)
  - \Sigma^{(1)} (\vare_a)\,,
\end{align}
whereas the vertex operator needs the subtraction of the first term only,
\begin{align}\label{gen:7}
\Lambda_{\rm vr} (\vare_a) \to  \Lambda_{\rm vr}^{(1+)} (\vare_a) =
  \Lambda_{\rm vr} (\vare_a) - \Lambda_{\rm vr}^{(0)} (\vare_a)\,,
\end{align}
where the superscript indicates the number of interactions with the binding field in the electron
propagator(s) inside the radiative photon loop. The double-vertex operator $\Lambda_{\rm dvr}$
contains three electron propagators inside the loop and thus is UV finite.

The separated terms containing zero and one interaction with the binding field ($\Sigma^{(0)}$,
$\Sigma^{(1)}$, $\Lambda_{\rm vr}^{(0)}$) are regularized by using the dimensional regularization
and calculated in momentum space. Their calculation does not involve bound electron propagators
and thus is beyond the scope of the present paper; we refer the reader for the original
investigations, Ref.~\cite{yerokhin:99:pra} for the self-energy matrix element, and
Ref.~\cite{yerokhin:04} for the magnetic vertex matrix element.

\section{Angular integration}

The integration over the angular variables in the above formulas is conveniently carried out with
help of the following representation of the matrix elements of the electron-electron interaction
operator,
\begin{equation} \label{Rdef}
 \lbr ab| I(\omega)| cd \rbr = \alpha\,
    \sum_{L = L_{\rm min}}^{L_{\rm max}} J_L(abcd)\, R_L(\omega, abcd)\,,
\end{equation}
where $J_L$ contains all the dependence on the angular momenta projections, $R_L$ are the radial
integrals defined in Appendix~\ref{sec:RJ}, and the summation over $L$ goes from $L_{\rm min} =
\max(|j_a-j_c|,|j_b-j_d|)$ to $L_{\rm max} = \min(j_a+j_c,j_b+j_d)$, with $j_n$ being the total
angular momentum of the electron state $n$. The function $J_L$ is given by
\begin{eqnarray}
  J_L(abcd)
  &=& \sum_{m_L} \frac{(-1)^{L-m_L+j_c-\mu_c+j_d-\mu_d}}{2L+1}\,
   \nonumber \\ && \times
  \,
  C^{Lm_L}_{j_a\mu_a,j_c-\mu_c}\, C^{Lm_L}_{j_d\mu_d,j_b-\mu_b}\,,
\end{eqnarray}
where $C_{j_1\mu_1,j_2\mu_2}^{j\mu}$ denotes the Clebsch-Gordan coefficient and $\mu_n$ is the
angular momentum projection of the electron state $n$.

Substituting Eq.~(\ref{Rdef}) into Eq.~(\ref{gen:1}) and performing the sum of two Clebsch-Gordan
coefficients, we immediately obtain the result for the matrix element of the self-energy
operator,
\begin{align} \label{eq:gr1}
\lbr a | \Sigma(\vare_a)| a\rbr =&\  \frac{i\alpha}{2\pi}\intinf d\omega
 \nonumber \\ & \times
\sum_{n,L}
 \, \frac{(-1)^{j_a-j_n+L}}{2j_a+1}
  \frac{R_L(\omega,anna)}{\vare_a -\omega - u\vare_n}\,.
\end{align}

In order to perform the angular integrations in the magnetic vertex operator, we first apply the
Wigner-Eckart theorem to the matrix element of the magnetic interaction $V_g$ (which is the
rank-1 spherical tensor),
\begin{align}
\lbr n_1 |V_g|n_2\rbr = \frac{(-1)^{j_1-\mu_1}}{\sqrt{3}}\,C^{10}_{j_2-\mu_2,j_1\mu_1}
\, ( n_1 || V_g || n_2 )\,,
\end{align}
where $(.||.||.)$ denotes the reduced matrix element. Now we can perform the angular integration
in the magnetic vertex matrix element as
\begin{align}\label{eq:gr2}
\sum_{\mu_{1}\mu_{2}}
\lbr an_2| I(\omega)|n_1a\rbr &\, \lbr n_1|V_{g}|n_2\rbr
= \sum_{L} X_L\,
 \nonumber \\ &
 \times
R_L(\omega,an_2n_1a)\,( n_1||V_{g}||n_2)
\,,
\end{align}
where $\mu_1$ and $\mu_2$ are the angular momentum projections of the electron states $n_1$ and
$n_2$, respectively, and $X_L$ are the angular coefficients defined by
\begin{align}
X_L = \sum_{\mu_1\mu_2}
\frac{(-1)^{j_1-\mu_1}}{\sqrt{3}}\,C^{10}_{j_2-\mu_2,j_1\mu_1}\,
 J_L(an_2n_1a)\,.
\end{align}
Performing the summation of three Clebsch-Gordan coefficients with help of formulas from
Ref.~\cite{varshalovich}, we obtain
\begin{align}
X_L = \frac{(-1)^{j_1-j_2}\,\mu_a}{\sqrt{j_a(j_a+1)(2j_a+1)}}
 \, \SixJ{j_1}{j_2}{1}{j_a}{j_a}{L}\,,
\end{align}
where $\{\ldots\}$ denotes the $6j$-symbol.

Analogously, performing summations of four Clebsch-Gordan coefficients with help of formulas from
Ref.~\cite{varshalovich}, we perform the angular integration for the double vertex matrix
elements,
\begin{align}\label{eq:gr3}
\sum_{\mu_1\mu_2\mu_3} &\,
 \lbr an_3| I(\omega)|n_1a\rbr\,
   \lbr n_1|V_{g}|n_2\rbr\,\lbr n_2|V_{\rm hfs}|n_3\rbr
=
 \nonumber \\ &
\sum_{L}
Z_L\, R_L(\omega, an_3n_1a)\,( n_1||V_{g}||n_2)\,( n_2||V_{\rm hfs}||n_3)
\,,
\end{align}
where the angular coefficients $Z_L$ are
\begin{align}
Z_L = \sum_{j_k} (-1)^{j_k+j_a}\,\Big[ C_{j_a\mu_a,10}^{j_k\mu_a}\Big]^2
 \, \NineJ{1}{j_1}{j_2}{j_a}{L}{j_3}{j_k}{j_a}{1}\,,
\end{align}
where $\{\ldots\}$ denotes the $9j$-symbol. In practical calculations, summations over the
angular momentum projections can be just as well carried out numerically.

The formulas above are written in terms of explicit summations over the Dirac spectrum, assuming
the spectral representation of the radial Green function. In order to use the analytical
representation of the Green function  in terms of regular and irregular solutions, we would need
to rewrite these formulas, identifying the components of the radial Green function,  as
\begin{align}
\sum_{n} \frac{g_{\kappa,n}(x)\,g_{\kappa,n}(x')}{E-\vare_n} \to G^{11}_{\kappa}(E,x,x')\,,
\ \ etc.
\end{align}
This is possible but often leads to long and unnecessary cumbersome expressions, especially for
complicated diagrams with multiple radial integrations. One can avoid this tedious work by
introducing \cite{yerokhin:03:epjd} the following formal representation for the radial Green
function
\begin{equation}\label{gr10}
G_{\kappa}(E,x,x') = \psi_{\kappa}(E,x)\,
        \tau^{T}_{\kappa}(E,x')\,,
\end{equation}
where the two-component functions $\psi_{\kappa}$ and $\tau_{\kappa}$ depend on {\em one} radial
argument only. The price to pay is that $\psi_{\kappa}$ and $\tau_{\kappa}$ have different forms
depending on the ordering of the radial arguments $x$ and $x'$,
\begin{equation}\label{gr1a}
\psi_{\kappa}(E,x) = \frac1{\Delta_{\kappa}^{1/2}} \times
  \left\{
        {\phi^0_{\,\kappa}(E,x),\ \mbox{\rm when }\  x<x'\,,}
        \atop
        {\phi^{\,\infty}_{\kappa}(E,x),\ \mbox{\rm when }\  x>x'\,,}
        \right.
\end{equation}
and
\begin{equation} \label{gr1b}
\tau_{\kappa}(E,x') = \frac1{\Delta_{\kappa}^{1/2}} \times
\left\{
        {\phi^{\,\infty}_{\kappa}(E,x'),\ \mbox{\rm when }\  x<x'\,,}
        \atop
        {\phi^{\,0}_{\kappa}(E,x'),\ \mbox{\rm when }\  x>x'\,.}
        \right.
\end{equation}
Here, $\phi^0_{\kappa}$ and $\phi^{\infty}_{\kappa}$ are the two-component solutions of the
radial Dirac equation bounded at the origin and infinity, respectively, and $\Delta_{\kappa}$ is
their Wronskian, see Eqs.~(\ref{g4}) and (\ref{g5}). Employing the representation (\ref{gr10}),
we can immediately use formulas written via summations over the Dirac spectrum for calculations
with the analytical representation of the Green function. The only complication is that the Green
function is discontinuous when the two radial arguments are equal, $x = x'$. This implies that
radial integrations in different matrix elements cannot be performed independently. Their
computation requires a special procedure, described in Sec.~\ref{sec:radial}.

\section{Choice of the integration contour}
\label{sec:CLH}

The formulas presented so far contained the integration over the virtual-photon energy $\omega$
performed along the real axis. This choice of the integration contour, however, is not favorable
for numerical calculations, since the Dirac Green function is a highly oscillating function for
large and real energy arguments and $x,x'\to \infty$. It is advantageous to deform the
integration contour to the region of large imaginary $\omega$ since the Dirac Green function
acquires an exponentially damping factor in this case. Deforming the contour of integration, one
should take care about poles and branch cuts of the integrand, however.

\begin{figure}
\centerline{\resizebox{0.5\textwidth}{!}{\includegraphics{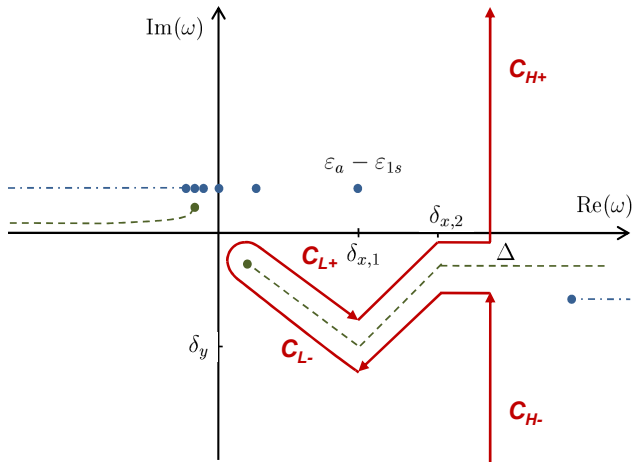}}}
 \caption{The poles and the branch cuts of the integrand of the matrix element of the self-energy
 operator and the integration contour $C_{LH}$ in the complex $\omega$ plane. The dashed lines
 (green) show the branch cuts of the photon
 propagator. The poles and the branch cuts of the electron propagator
 are shown by dots and the dashed-dot line (blue). The solid line (red) shows the integration
 contour $C_{LH}$.  \label{fig:CLH}}
\end{figure}

The analytical structure of the Dirac Green function is outlined in Sec.~\ref{sec:green}. The
branch cuts of the photon propagator (\ref{g3}) are defined by the square root function, which
should be understood as a limit of the regularized expression with a photon mass $\mu>0$,
\begin{align}
\sqrt{\omega^2 + i0} &\to \lim_{\mu\to 0} \sqrt{\omega^2-\mu^2+i0}
 \nonumber \\
 &= \lim_{\mu\to 0} \sqrt{\omega-\mu+i0}\,\sqrt{\omega+\mu-i0}\,.
\end{align}
The photon propagator thus has two branch cuts starting from $\omega = \mu - i0$ and $\omega =
-\mu + i0$. The analytical structure of the integrand of the self-energy matrix element is shown
in Fig.~\ref{fig:CLH}.

Fig.~\ref{fig:CLH} also presents the deformed contour of the $\omega$ integration, which we found
to be optimal for most practical calculations. Specifically, the contour $C_{LH}$ consists of the
low-energy part $C_L$ and the high-energy part $C_H$.

The low-energy part of the integration contour $C_L$ consists of two parts, $C_{L+}$ and
$C_{L-}$, the first of which runs on the upper bank of the cut of the photon propagator and the
second, on the lower bank and in the opposite direction. On the upper bank $\sqrt{\omega^2} =
\omega$, whereas on the lower bank $\sqrt{\omega^2} = -\omega$. The integrands for $C_{L+}$ and
$C_{L-}$ differ only by the sign of $\omega$ in the photon propagator (and the overall sign due
to the opposite directions of the integration), thus allowing the following simplification,
$$
e^{i\omega\,x_{12}} \to e^{i\omega\,x_{12}} - e^{-i\omega\,x_{12}} = 2\,i\, \sin(\omega\,x_{12}).
$$
The high-energy part $C_H$ is parallel to the imaginary axis and consists of two parts $C_{H-} =
(\Delta-i\infty, \Delta-i\epsilon)$ and from $C_{H+} = (\Delta+i\epsilon, \Delta+i\infty)$. The
integrands for $C_{H+}$ and $C_{H-}$ are typically complex conjugated, so that one can perform
the integration over $C_{H+}$ only, take the real part of the result and multiply by two.

In the general case of an excited reference state, the low-energy part $C_L$ is bent in the
complex plane, in order to avoid singularities coming from virtual bound states with energies
$\vare_n < \vare_a$ in the electron propagator. Specifically, the contours $C_{L+}$ and $C_{L-}$
consist of 3 sections: $(0,\delta_{x,1}-i\delta_y)$, $(\delta_{x,1}-i\delta_y,\delta_{x,2})$, and
$(\delta_{x,2},\Delta)$, as shown on Fig.~\ref{fig:CLH}.  The parameters of the contour
$\delta_{x,1}$, $\delta_{x,2}$, $\delta_{y}$, and $\Delta$ may be chosen differently. In our
calculations, we used the following choice (assuming the reference state $a$ to be an excited
state):  $\Delta = \Za\,\vare_a$; $\delta_{x,1} = \vare_a - \vare_{1s}$; $\delta_{x,2} =
2\,\delta_{x,1}$; $\delta_y = \delta_{x,1}/2$. If the reference state $a$ is the ground state,
there is no need to bend the low-energy part of the contour in the complex plane (as there are no
intermediate states with energy $0 <\vare_n < \vare_a$); so we just integrate along the real axis
(setting $\delta_y = 0$).

We note that the described contour $C_{LH}$ resembles the contour used by P.~Mohr in his
calculations \cite{mohr:74:a}. The difference is that he did not bend the low-energy part in the
complex plane and used a different choice of the parameter $\Delta$, $\Delta = \vare_a$.

Another choice of the $\omega$ integration contour frequently encountered in the literature
(e.g., in Refs.~\cite{blundell:91:se,sapirstein:01:lamb,sapirstein:03:hfs}) is the standard Wick
rotation from the real into the imaginary axis, $\omega \to i\omega$. In this case the
intermediate states with energy $0 <\vare_n \le \vare_a$ lead to appearance of the pole terms,
which need a special treatment. Apart for the pole contributions, small energy differences
$\vare_a-\vare_n$ appear in the denominators of the electron propagators, leading to a rapidly
varying structure of the integrand for small $\omega$ in this choice of the contour, which may
lead to numerical difficulties.

\section{Infrared divergencies}

In this section we address the infrared (IR) reference-state divergencies which appear in the QED
corrections involving bound-electron propagators.

We start with pointing out that the bound-state QED corrections do not possess the standard
free-QED IR divergences which arise when the four-momentum $p$  of the intermediate electron
states approaches the mass shell, $\rho = (m^2-p^2)/m^2 = (m^2-p_0^{2}+\bfp^2)/m^2 \to 0$. For
the bound-state QED corrections, the intermediate electron states are always off mass shell ($p_0
= \vare_a < m \Rightarrow \rho > 0$) and the would-be IR divergences are cut off by the binding
energy of the reference state. However, the bound-state QED corrections often contain  IR
divergences of a different kind, also known as the reference-state divergences. They appear when
two or more denominators in the electron propagators vanish at $\omega \to 0$. Specifically, IR
divergences arise in the magnetic vertex operator (\ref{gen:4}) when $n_1 = n_2 = a$ and in the
double vertex operator (\ref{gen:5}) when $n_1 = n_3 = a$.

The general approach to the treatment of the IR divergencies is to separate out the divergent
contributions, regularize them by introducing a finite photon mass $\mu$ in the photon
propagator, evaluate the integral over $\omega$ analytically, and separate out the
$\mu$-dependent divergent terms. The divergent terms should of course cancel out when all
relevant contributions are summed together. The evaluation of the IR-divergent integrals with the
finite photon mass is illustrated in Appendix~\ref{app:IR}.

\begin{widetext}
Using formulas from Appendix~\ref{app:IR}, the magnetic vertex matrix element (\ref{gen:4}) can
be transformed to a form that is explicitly free from any IR divergences,
\begin{align} \label{gen:4:ir}
\lbr a | \Lambda_{\rm vr}(\vare_a)| a\rbr
= &\,
\frac{i}{2\pi}
  \intinf d\omega \,  \Bigg[
\sum_{n_1n_2}
\frac{\lbr an_2| I(\omega)|n_1a\rbr\,\big[\lbr n_1|V_{g}|n_2\rbr - \delta_{n_1n_2}\, \lbr a|V_{g}|a\rbr\big]}
  {(\vare_a-\omega-u\,\vare_{n_1})(\vare_a-\omega-u\,\vare_{n_2})}
  \nonumber \\ &
- \sum_{\mu_{a'}\mu_{a''}}
\frac{\lbr aa''| I(\omega)|a'a\rbr\,\big[\lbr a'|V_{g}|a''\rbr - \delta_{a'a''} \, \lbr a|V_{g}|a\rbr\big]}
  {(-\omega+i0)^2}\Bigg]
  \nonumber \\ &
  + \frac{\alpha}{\pi}
  \sum_{\mu_{a'}\mu_{a''}}
   \lbr aa''| \alpha_{\mu}\alpha^{\mu}\,\ln x_{12}|a'a\rbr\,\big[\lbr a'|V_{g}|a''\rbr
     - \delta_{a'a''}\,\lbr a|V_{g}|a\rbr\big]
\,,
\end{align}
where $a'$ and $a''$ denote the reference state $a$ with a different momentum angular projection
($\mu_{a'}$ and $\mu_{a''}$, respectively).
\end{widetext}

For the double-vertex matrix element (\ref{gen:5}) the situation is somewhat more complicated
because there are two types of divergences, the one $\propto 1/\mu$ coming from three vanishing
denominators ($n_1 = n_2 = n_3 = a$) and the other $\propto \ln \mu$ coming from two vanishing
denominators ($n_1 = n_3  = a \neq n_2$). Still, the divergences in the double-vertex matrix can
be handled with help of formulas from Appendix~\ref{app:IR} analogously to that for the magnetic
vertex case.

There exists also a more economic method of handling IR divergences in actual calculations. It
relies on the fact that the matrix elements (\ref{gen:4}) and (\ref{gen:5}) are defined so that
they are overall IR finite, i.e., have a well-defined limit at $\mu\to 0$. This means that they
can be numerically evaluated with the zero photon mass. As long as the $\omega$ integration is
performed {\em after} all parts of the integrand are combined together, the integrand will have a
smooth small-$\omega$ behaviour when integrated along the low-energy part of the integration
contour $C_{LH}$. The would-be IR divergences will be cancelled numerically at a given $\omega$
between different parts of the integrand. It is easy to check that for the magnetic vertex matrix
element, both methods give the identical numerical results. For the double vertex matrix element,
the numerical treatment of IR divergences was used in the calculation of the diamagnetic
shielding in Refs.~\cite{yerokhin:11:prl,yerokhin:12:shield}.

It should be mentioned that vanishing denominators in the electron propagators could arise not
only from the intermediate states $n = a$, but also from the intermediate states having the same
energy but the opposite parity as the reference state (e.g., $n = 2p_{1/2}$ for $a = 2s$ for the
point nuclear model). Such intermediate states do not cause IR divergences, since the radial
matrix element in the numerator vanishes due to the orthogonality of the wave functions, as can
be seen from formulas in Appendix~\ref{app:IR}.

A different approach for handling the IR divergencies was used by the Notre-Dame group
\cite{blundell:97:pra,sapirstein:01:lamb,sapirstein:01:hfs}. In their works, numerical
calculations were performed with an explicit regularization parameter $\delta$ shifting the
position of the reference-state energy, $\vare_a \to \vare_a(1-\delta)$; the numerical limit of
$\delta \to 0$ was then performed in the end of the calculations.

\section{Radial integration}
\label{sec:radial}

In actual calculations it is very important to find an efficient way to perform multiple radial
integrations. The number of radial integrations is two for the self-energy matrix element, three
for the magnetic vertex, and four for the double-vertex matrix element. In what follows we will
assume that the analytical representation of the Dirac Green function is used, since in the
basis-set representation, the radial integrations do not cause particular difficulties.

We now formulate a general numerical approach suitable for carrying out multiple radial
integrations, first introduced in the context of the two-loop self-energy in
Ref.~\cite{yerokhin:03:epjd}. We start with the simplest case of the self-energy matrix element,
in which the radial integral is two-dimensional. The two-dimensional radial integrals can be
schematically represented to be a linear combination of terms of the following structure
\begin{equation} \label{integral}
\int_0^{\infty}dx_1\, \int_0^{\infty}dx_2\, H(x_1)\,I(x_<)\, L(x_>)\,M(x_2) \,,
\end{equation}
where  $x_> = \max(x_1,x_2)$ and $x_< = \min(x_1,x_2)$ and $H$, $I$, $L$, $M$ are some functions
of the specified arguments. It is important that the integrand can be represented as a product of
functions that depend on one radial argument only, some of those being $x_<$ and $x_>$. In
particular, $I(x_<)$ involves $\phi^0(x_<)$ from the Dirac Green function and $j_l(\omega x_{<})$
from the photon propagator, and $L(x_>)$ involves $\phi^{\infty}(x_>)$ and $h^{(1)}_l(\omega
x_{>})$. It is clear that if we store all functions on a suitably chosen radial grid, it should
be possible to compute the integral (\ref{integral}) just by summing up the pre-stored data.

In order to do this, we introduce a 3-dimensional radial grid $\left\{r_{i,j,k}\right\}$ on the
interval $(0,r_{\rm max})$ as follows. First, we fill the elements of the first layer,
$r_{i,0,0}$ with $i = 0,\ldots,N_i$, which coarsely span the whole interval, {\em e.g.}, as
\begin{equation}
r_{i,0,0} = r_0\,\frac{1-t^2}{t^2}\,,
\end{equation}
where $t$ is uniformly distributed on the interval $(t_{\rm min},1)$ and $t_{\rm min}\approx 0$
is defined by the cavity radius $r_{\rm max}$. Next, we introduce a finer grid of the second
layer. Specifically, on each interval $(r_{i,0,0},r_{i+1,0,0})$ we introduce the set of the
Gauss-Legendre abscissae $\left\{r_{i,j,0}\right\}_{j=1}^{N_j}$. We see that in order to perform
the outer radial integration, it is sufficient to know the integrand on the grid
$\left\{r_{i,j,0}\right\}$.

In order to perform the inner radial integral, we have to split the integration interval at the
point $x_2 = x_1$, since it is the discontinuity point of the integrand. We achieve this by
introducing a yet finer grid of the third layer, $\left\{r_{i,j,k}\right\}$. Specifically, for
fixed values of $i$ and $j$, the set $\left\{r_{i,j,k}\right\}_{k = 1}^{N_k}$ represents the
Gauss-Legendre abscissae on the interval $(r_{i,j,0},r_{i,j+1,0})$ if $j<N_j$ and on the interval
$(r_{i,j,0},r_{i+1,0,0})$ if $j = N_j$. Now, for each point $r_{i,j,0}$ of the outer radial
integration, we can perform the inner integral splitted into parts, $(0,r_{i,j,0})$ and
$(r_{i,j,0},r_{\rm max})$.

We conclude that when all functions in the integrand of Eq.~(\ref{integral}) are stored on the
radial grid $\left\{r_{i,j,k}\right\}$, the two-dimensional numerical integration can be carried
out by just summing up the pre-stored numerical values. The described procedure can be easily
generalized for integrals of higher dimensions. So, for a computation of a four-dimensional
radial integral, we need a 5-dimensional grid $\left\{r_{i,j,k,l,m}\right\}$ introduced in the
same way as $\left\{r_{i,j,k}\right\}$.

An additional complication arises from the fact that the regular Dirac solution
$\phi^0_{\kappa}(\vare,r)$ in the Dirac Green function has typically the exponentially-growing
behaviour for large values of the radial argument and complex values of $\vare$, whereas the
irregular solutions $\phi^{\infty}_{\kappa}(\vare,r)$ is exponentially decreasing in this region.
In order to avoid numerical overflow and underflow, we store the ``normalized'' solutions
$\widetilde{\phi}^0$ and $\widetilde{\phi}^{\infty}$, with the approximate large-$r$ and
small-$r$ behaviour pulled out,
\begin{align} \label{eq:norm1}
\phi^0_{\kappa}(\vare,r) &\, = r^{|\kappa|}\,e^{cr}\, \widetilde{\phi}^0_{\kappa}(\vare,r) \,, \\
\phi^{\infty}_{\kappa}(\vare,r) &\, = r^{-|\kappa|}\,e^{-cr}\, \widetilde{\phi}^{\infty}_{\kappa}(\vare,r) \,,
\label{eq:norm2}
\end{align}
where $c = \sqrt{1-(\vare/m)^2}$\,. When $\phi^0_{\kappa}(\vare,r)$ and
$\phi^{\infty}_{\kappa}(\vare,r)$ multiply together in the Dirac Green function, the result is
usually in the range accessible in the standard double-precision (8-byte) arithmetics. A similar
normalization is required also for the regular $j_l$ and irregular $h^{(1)}_l$ Bessel solutions,
originating from the photon propagator. With these precautions, we are able to perform
calculations completely within the standard double-precision arithmetics typically for $\kappa
\le 50$. For $\kappa \le 100$, it is usually possible to use the quadruple-precision arithmetics
for computation of Dirac ($\phi^0_{\kappa}$, $\phi^{\infty}_{\kappa}$) and Bessel ($j_l$,
$h^{(1)}_l$) solutions but the double-precision arithmetics for the radial integrations. For even
higher values of $\kappa$, use of the extended-precision arithmetics becomes unavoidable
\cite{yerokhin:17:pra:segfact}.

\section{Magnetically-perturbed Green function}
\label{sec:Gmagn}

The computation of radial integrations in diagrams with various kind of potentials can be
significantly accelerated by introducing the first-order perturbations of the Green function by
this potentials. Such an approach was used long ago by Gyulassy in his evaluation of the vacuum
polarization \cite{gyulassy:75}. More recently, similar algorithms were used in calculations of
various self-energy corrections (in particular, in
Refs.~\cite{artemyev:07:prl,yerokhin:11:fns,artemyev:13}).

In this section we describe the computation of the Dirac Green function perturbed by a magnetic
potential $V_g$, which will be referred to as the magnetically-perturbed Green function.
Specifically, we are interested in the radial part of the magnetically-perturbed Green function,
defined as
\begin{align}
{\cal G}_{\kappa_1\kappa_2}(x_1,x_3) =
   \int_0^{\infty}dx_2\,x_2^2\, G_{\kappa_1}(x_1,x_2)\,V_g(x_2)\,G_{\kappa_2}(x_2,x_3)\,,
\end{align}
where $V_g(x) = x\,\sigma_x$ is the radial part of the magnetic potential $V_g(\bfx)$. Using the
representation of the radial Green functions in terms of the regular and irregular Dirac
solutions, see Eq.~(\ref{g4}), we obtain the following expressions for the magnetically-perturbed
Green function. For $x_1 < x_3$, we get
\begin{align}
{\cal G}_{\kappa_1 \kappa_2}&(x_1,x_3) =  \phi^{\infty}_{\kappa_1}(x_1)\,
                                          \Phi^{0\,0}_{\kappa_1\kappa_2}(x_1)
                                             \phi^{\infty^T}_{\kappa_2}(x_3)
 \nonumber \\ &
 + \phi^{0}_{\kappa_1}(x_1)\,
                                          \Phi^{\infty\,\infty}_{\kappa_1\kappa_2}(x_3)
                                             \phi^{0^T}_{\kappa_2}(x_3)\,,
 \nonumber \\ &
   +
    \phi^{0}_{\kappa_1}(x_1)\,\Big[\Phi^{\infty\, 0}_{\kappa_1\kappa_2}(x_3)
                                - \Phi^{\infty\, 0}_{\kappa_1\kappa_2}(x_1)\Big]\,
                                             \phi^{\infty^T}_{\kappa_2}(x_3)
                                             \,,
\end{align}
whereas for $x_1 > x_3$,
\begin{align}
{\cal G}_{\kappa_1 \kappa_2}&(x_1,x_3) =  \phi^{\infty}_{\kappa_1}(x_1)\,
                                          \Phi^{0\,0}_{\kappa_1\kappa_2}(x_3)
                                             \phi^{\infty^T}_{\kappa_2}(x_3)
 \nonumber \\ &
                                   + \phi^{0}_{\kappa_1}(x_1)\,
                                          \Phi^{\infty\,\infty}_{\kappa_1\kappa_2}(x_1)
                                             \phi^{0^T}_{\kappa_2}(x_3)\,,
 \nonumber \\ &
   +
    \phi^{\infty}_{\kappa_1}(x_1)\,\Big[\Phi^{0\,\infty}_{\kappa_1\kappa_2}(x_1)
                                - \Phi^{0,\infty}_{\kappa_1\kappa_2}(x_3)\Big]\,
                                             \phi^{0^T}_{\kappa_2}(x_3)
                                             \,,
\end{align}
where for simplicity we assumed that $\phi^{0}_{\kappa}$ and $\phi^{\infty}_{\kappa}$ are
normalized so that their Wronskian is unity and the functions $\Phi_{\kappa_1\kappa_2}$ are
defined by the integrals
\begin{align}
\Phi^{0\,0}_{\kappa_1\kappa_2}(x) = &\, \int_0^{x}dx_2\,x_2^2\,
        \phi^{0^T}_{\kappa_1}(x_2)\,
               V_g(x_2)\,
            \phi^{0}_{\kappa_2}(x_2)\,,
            \\
\Phi^{0\,\infty}_{\kappa_1\kappa_2}(x) = &\, \int_0^{x}dx_2\,x_2^2\,
        \phi^{0^T}_{\kappa_1}(x_2)\,
               V_g(x_2)\,
            \phi^{\infty}_{\kappa_2}(x_2)\,,
            \\
\Phi^{\infty\,0}_{\kappa_1\kappa_2}(x) = &\, \int_0^{x}dx_2\,x_2^2\,
        \phi^{\infty^T}_{\kappa_1}(x_2)\,
               V_g(x_2)\,
            \phi^{0}_{\kappa_2}(x_2)\,,
            \\
\Phi^{\infty\,\infty}_{\kappa_1\kappa_2}(x) = &\, \int_{x}^{\infty}dx_2\,x_2^2\,
        \phi^{\infty^T}_{\kappa_1}(x_2)\,
               V_g(x_2)\,
            \phi^{\infty}_{\kappa_2}(x_2)\,.
\end{align}

We observe that after storing the functions $\phi_{\kappa}(x)$ and $\Phi_{\kappa_1,\kappa_2}(x)$
on a radial grid, we are able to construct the magnetically-perturbed Green function ${\cal
G}_{\kappa_1\kappa_2}(x_1,x_2)$ for any radial arguments needed in our computation. The integral
functions $\Phi_{\kappa_1\kappa_2}(x)$ are evaluated by numerical integration with help of
Gauss-Legendre quadratures. It is important that only one integral over $(0,\infty)$ needs to be
evaluated (for a given value of the energy argument) in order to store
$\Phi_{\kappa_1\kappa_2}(x)$ on the whole radial grid. Analogously to the case of the plain Green
function, all manipulations with the regular and irregular solutions need to be carried out after
normalizing them according to Eqs.~(\ref{eq:norm1}) and (\ref{eq:norm2}) in order to prevent
numerical overflow.

\section{Numerical calculations}
\label{sec:numerics}

In this section we demonstrate the technique described in previous sections with three examples
of actual calculations. The first one is the calculation of the one-loop self-energy correction
to the Lamb shift of a hydrogen-like ion. In Table~\ref{tab:se} we present numerical results for
the one-loop self-energy correction to the Lamb shift of the $2s$ state of hydrogen-like calcium
($Z = 20$), for the point nuclear model.

The many-potential part $\lbr \Sigma^{(2+)}\rbr$ defined by Eq.~(\ref{gen:6}) is calculated in
coordinate space by the method described in the present work. Specifically, the one-potential
Green function was calculated by the method described in Sec.~\ref{sec:Gmagn}. (Alternatively, it
can also be calculated as a derivative over the nuclear charge $Z$, as described in
Ref.~\cite{yerokhin:99:pra}.) For the radial integration, we used a four-dimensional grid $\big\{
r_{i,j,k,l}\big\}$ constructed as discussed in Sec.~\ref{sec:radial} with the number of
integration points $(N_i,N_j,N_k,N_l) = (15,10,6,6)$. The $\omega$ integration is carried out
along the contour $C_{LH}$ introduced in Sec.~\ref{sec:CLH} using the Gauss-Legendre quadratures,
after mapping of the integration intervals to the range $(0,1)$. The summation over the partial
waves was extended up to $|\kappa| = 60$, with the remaining tail of the expansion estimated by a
least-square fitting in polynomials in $1/|\kappa|$. The remaining zero- and one-potential part
$\lbr \Sigma^{(0+1)}\rbr$ is calculated in momentum space. Their computation is relatively simple
and can be performed up to essentially arbitrary accuracy. This part  is not discussed here; we
refer the reader to the original work \cite{yerokhin:99:pra}.

As follows from Table~\ref{tab:se}, the uncertainty of the final numerical result for the
self-energy correction comes exclusively from the truncation of the partial-wave expansion. It
can be seen that despite the inclusion of 60 partial waves, the resulting accuracy is
significantly lower than that of the best literature values. There are two ways described in the
literature that allow to achieve a better numerical precision. One method was developed
originally by Mohr \cite{mohr:74:a,mohr:74:b,mohr:92:b} and extended by Jentschura and Mohr
\cite{jentschura:99:prl,jentschura:01:pra}. This method involves a summation of many thousands of
partial waves and usage of extended-precision arithmetics in order to obtain very accurate
numerical results. Another method was developed in Ref.~\cite{yerokhin:05:se}. It involves an
additional subtraction in $\Sigma^{(2+)}$ which greatly accelerates the convergence of the
partial-wave expansion and allows one to obtain accurate numerical results with just 20-30
partial waves. Both these methods are difficult to extend for computations of more complicated
diagrams, unfortunately.

Table~\ref{tab:vermag} presents our numerical results for the self-energy correction to the $g$
factor of the $2s$ state of hydrogen-like calcium ($Z = 20$), for the point nuclear model. The
many-potential part $\lbr \Lambda_{\rm vr}^{(2+)}\rbr$ is calculated in coordinate space by the
method described in the present work. It is important that we calculate the magnetic vertex after
subtracting  {\em two} first terms of its potential expansion, not just one as in
Eq.~(\ref{gen:7}). This is done in order to accelerate the convergence of the partial-wave
expansion of the matrix element, following Refs.~\cite{persson:97:g,yerokhin:04}. The subtracted
part $\lbr \Lambda_{\rm vr}^{(0+1)}\rbr$ is calculated in momentum space as described in
Ref.~\cite{yerokhin:04}. The irreducible part $\lbr \Lambda_{\rm ir}\rbr$ is expressed as a
non-diagonal matrix element of the self-energy operator; its numerical values were taken from
Ref.~\cite{yerokhin:04}. The total result presented in Table~\ref{tab:vermag} is in good
agreement with the previous value obtained in Ref.~\cite{yerokhin:04}. Its numerical uncertainty
comes exclusively from the truncation of the partial-wave expansion. Even more accurate results
can be achieved if one extends the partial-wave expansion further, as was done for the $1s$ state
in Ref.~\cite{yerokhin:17:pra:segfact}, but it requires significant efforts and intensive usage
of extended-precision arithmetics.

In Table~\ref{tab:shield} we present numerical results for the self-energy correction to the
diamagnetic shielding constant of the $1s$ state of hydrogen-like calcium ($Z = 20$), for the
point nuclear model. The many-potential part $\lbr \Lambda_{\rm dvr}\rbr$ is calculated in
coordinate space by the method described in the present work. We observe a slow convergence of
the partial-wave expansion of the results presented in the table. It can probably be accelerated
by separating out the leading term of the potential expansion (i.e., the contribution of the free
propagators) and calculating it in the momentum space, but this has not been accomplished so far.
The other contributions to the shielding constant are defined in Ref.~\cite{yerokhin:12:shield};
the corresponding numerical results are taken from that work. Again, we observe that the dominant
uncertainty of the final result comes from the truncation of the partial-wave expansion.

\section{Summary}

In this paper we described the technique used in modern calculations of QED corrections with the
bound-electron propagators, including the notoriously problematic diagrams with several
propagators inside the radiative photon loop. The bound-electron propagators are described by the
Green function of the Dirac equation with the binding nuclear potential. We considered two most
widely used ways to represent the Dirac Green function, the representation via the regular and
irregular Dirac solutions and the finite basis set representation. These representations are
applicable for a wide range of binding potentials, including the case of the nuclear field
modified by a spherically-symmetric screening potential caused by the presence of other electrons
in the atom.

We demonstrated that the dominant uncertainty of the obtained results usually comes from the
truncation of the partial-wave expansion. Further extension of the partial-wave expansion is
possible but often associated with large technical difficulties. In view of this, it is important
to look for ways to accelerate convergence of the partial-wave expansion. This was accomplished
for the one-loop self-energy in Ref.~\cite{yerokhin:05:se}, for the self-energy correction to the
$g$ factor in Refs.~\cite{yerokhin:02:prl,yerokhin:04}, and for the self-energy correction to the
hyperfine splitting in Refs.~\cite{yerokhin:08:prl,yerokhin:10:sehfs}. Unfortunately, all these
methods turned out to be problem-specific, {\em i.e.}, they do not allow straightforward
extensions to more complicated corrections. It would be thus of great importance to find a more
universal approach to improve the convergence of the partial-wave expansion in such calculations.

\begin{acknowledgments}
The work presented in this paper was supported by the Russian Science Foundation (Grant No.
20-62-46006).
\end{acknowledgments}

\begin{table}[t]
\vspace*{0.75cm}
\caption{Numerical results for the one-loop self-energy correction for the $2s$ state of hydrogen-like calcium ($Z = 20$),
for the point nucleus, in terms of the standard scaled function
$F(\Za) = \delta E/[(\alpha/\pi)\,(\Za)^4/n^3]$,
where $\delta E$ is a contribution to the energy in relativistic units.
$S_{l}$ denotes the sum of partial-wave expansion $\sum_{|\kappa| = 1}^l$;
$\delta S_{l}$ is the increment with respect to the previous line.
\label{tab:se}}
\begin{ruledtabular}
\begin{tabular}{ll w{4.8}  w{4.8}  }
 & \multicolumn{1}{l}{$l$}
  & \multicolumn{1}{c}{$S_{l}$}
     & \multicolumn{1}{c}{$\delta S_{l}$}  \\
\hline\\[-5pt]
$\lbr \Sigma^{(2+)}\rbr$
 & 1  &  82.268\,19  &        \\
 & 2  &  85.541\,56  &   3.273\,37      \\
 & 3  &  86.515\,41  &   0.973\,85      \\
 & 4  &  86.967\,68  &   0.452\,26      \\
 & 5  &  87.223\,70  &   0.256\,02      \\
 &10  &  87.675\,13  &   0.451\,44     \\
 &15  &  87.790\,89  &   0.115\,75     \\
 &20  &  87.836\,31  &   0.045\,42     \\
 &30  &  87.869\,99  &   0.033\,68     \\
 &40  &  87.881\,51  &   0.011\,52     \\
 &50  &  87.886\,57  &   0.005\,06     \\
 &60  &  87.889\,17  &   0.002\,61    \\
 &$\infty$
     &  87.894\,34\,(26)
                   &    0.005\,17\,(26)  \\
$\lbr \Sigma^{(0+1)}\rbr$
     && -84.387\,704 \\
Total&&   3.506\,64\,(26) \\
P.~J.~Mohr \cite{mohr:92:b} &&
         3.506\,648\,(2) \\
Refs.~\cite{yerokhin:05:se,shabaev:13:qedmod} &&
         3.506\,647\,(5)\\
\end{tabular}
\end{ruledtabular}
\end{table}

\begin{table}[t]
\caption{Numerical results for the self-energy correction to the
$g$ factor of the $2s$ state of hydrogen-like calcium ($Z = 20$),
for the point nucleus (in units of $10^{-6}$).
\label{tab:vermag}}
\begin{ruledtabular}
\begin{tabular}{ll w{4.8}  w{4.8}  }
 & \multicolumn{1}{l}{$l$}
  & \multicolumn{1}{c}{$S_{l}$}
     & \multicolumn{1}{c}{$\delta S_{l}$}  \\
\hline\\[-5pt]
$\lbr \Lambda_{\rm vr}^{(2+)}\rbr$
&  1  &  36.130\,52  &               \\
&  2  &  17.563\,35  &   -18.567\,17 \\
&  3  &  14.605\,25  &    -2.958\,10 \\
&  4  &  13.586\,86  &    -1.018\,39 \\
&  5  &  13.115\,22  &    -0.471\,64 \\
& 10  &  12.489\,16  &    -0.626\,06 \\
& 15  &  12.379\,38  &    -0.109\,78 \\
& 20  &  12.343\,92  &    -0.035\,46 \\
& 25  &  12.328\,78  &    -0.015\,15 \\
& 30  &  12.321\,11  &    -0.007\,67 \\
& 35  &  12.316\,76  &    -0.004\,35 \\
&$\infty$ &
         12.306\,43\,(50)& -0.010\,33\,(50) \\
$\lbr \Lambda_{\rm vr}^{(0+1)}\rbr$
&&     2237.914\,11  &               \\
$\lbr \Lambda_{\rm ir}\rbr$
&&       75.453\,02  &               \\
Total&&2325.673\,56\,(50)            \\
Ref.~\cite{yerokhin:04}
     &&2325.674\,(5)\\
\end{tabular}
\end{ruledtabular}
\end{table}

\begin{table}[t]
\caption{Numerical results for the self-energy correction to the
magnetic shielding constant $\sigma$ of the $1s$ state of hydrogen-like calcium ($Z = 20$),
for the point nucleus, in units of the scaled function $D(\Za) =
\delta \sigma/[\alpha^2(\Za)^3]$ where $\delta\sigma$ is a contribution to the shielding
constant.
\label{tab:shield}}
\begin{ruledtabular}
\begin{tabular}{ll w{4.8}  w{4.8}  }
 & \multicolumn{1}{l}{$l$}
  & \multicolumn{1}{c}{$S_{l}$}
     & \multicolumn{1}{c}{$\delta S_{l}$}  \\
\hline\\[-5pt]
$\lbr \Lambda_{\rm dvr}\rbr$
&  1 &  -3.409\,2 &               \\
&  2 &  -5.550\,9 &  -2.141\,7    \\
&  3 &  -6.559\,8 &  -1.008\,9    \\
&  4 &  -7.111\,6 &  -0.551\,7    \\
&  5 &  -7.438\,5 &  -0.327\,0    \\
& 10 &  -7.941\,8 &  -0.503\,2    \\
& 15 &  -7.986\,1 &  -0.044\,3    \\
& 20 &  -7.968\,3 &   0.017\,7    \\
& 25 &  -7.945\,0 &   0.023\,3    \\
& 30 &  -7.925\,5 &   0.019\,5    \\
& 35 &  -7.910\,4 &   0.015\,1    \\
&  $\infty$
     & -7.846\,7\,(32) &   0.063\,7\,(32)   \\
$\lbr \Lambda_{\rm der}\rbr$
     &&  7.782\,4 \\
$\lbr \Lambda_{\rm vr, Zee}\rbr$
     &&  1.760\,7 \\
$\lbr \Lambda_{\rm vr, hfs}\rbr$
     && -0.404\,9 \\
$\lbr \Lambda_{\rm po}\rbr$
     && -2.217\,0 \\
Total&& -0.925\,5\,(32)&\\
\end{tabular}
\end{ruledtabular}
\end{table}

%
%
%

\appendix

\begin{widetext}
%
\section{Relativistic Slater radial integral} \label{sec:RJ}

The matrix element of the electron-electron interaction operator (\ref{gen:2}) is represented in
the form (\ref{Rdef}), where $R_J(\omega,abcd)$ is the relativistic generalization of the Slater
radial integral. The explicit expression for $R_J$ can be obtained, {\em e.g.}, by reformulating
formulas presented in Appendix of Ref.~\cite{johnson:88:b}. The result for the radial integral
$R_J$ in the Feynman gauge is written as \cite{yerokhin:99:pra}
\begin{align} \label{P1}
R_J(\omega,abcd) = (2J+1)\int_0^{\infty}dx_2\,dx_1\,(x_1x_2)^2\,
   & \Big[ (-1)^J C_J(\kappa_a,\kappa_c)\, C_J(\kappa_b,\kappa_d)\,
    g_J(\omega,x_<,x_>)\, W_{ac}(x_1)\, W_{bd}(x_2) \nonumber \\
&   - \sum_L (-1)^L g_L(\omega,x_<,x_>)\, X_{ac,JL}(x_1)\, X_{bd,JL}(x_2)
        \Big] \ ,
\end{align}
where $x_> = \max(x_1,x_2)$, $x_< = \min(x_1,x_2)$, the functions $W_{ab}$ and $X_{ab,JL}$ are
defined by
\begin{align} \label{P1}
W_{ab}(x) =&\, g_a(x)\,g_b(x)+f_a(x)\,f_b(x) \ , \\
X_{ab,JL}(x) =&\,  g_a(x)\,f_b(x)\, S_{JL}(-\kb,\ka)
                    -f_a(x)\,g_b(x)\, S_{JL}(\kb,-\ka)  \,.
\end{align}
Here, $g_n$, $f_n$ are the upper and the lower radial components of the Dirac wave function,
respectively. The function $g_l(\omega,x_<,x_>)$ is the radial part of the partial-wave expansion
of the photon propagator,
\begin{eqnarray} \label{eiomega}
\frac{e^{i\omega x_{12}}}{x_{12}} &=& \sum_{l} (2l+1)\,
        g_l(\omega, x_<, x_>)\, P_l(\xi) \ ,
\end{eqnarray}
where $P_l(\xi)$ is the Legendre polynomial, $\xi = \hx_1\cdot\hx_2$,
\begin{eqnarray}
g_l(0, x_<, x_>) &=& \frac1{2l+1} \frac{x_<^l}{x_>^{l+1}} \ ,\\
        \label{eiomega2}
g_l(\omega, x_<, x_>) &=& i\omega\, j_l(\omega x_<)
h^{(1)}_l(\omega x_>) \ ,
\end{eqnarray}
 and $j_l(z)$,
$h^{(1)}_l(z)$ are the spherical Bessel functions. The angular coefficients
$S_{JL}(\kappa_a,\kappa_b)$  differ from the zero only for $L = J-1$, $J$, $J+1$ and can be
written for $J \ne 0$ as follows:
\begin{eqnarray} \label{SJL}
S_{J\, J+1}(\ka ,\kb ) &=& \sqrt{\frac{J+1}{2J+1}}
 \left( 1+ \frac{\ka +\kb}{J+1} \right) C_J(-\kb ,\ka ) \ , \\
S_{J\, J}(\ka ,\kb ) &=& \frac{\ka -\kb}{\sqrt{J(J+1)}}
 C_J(\kb ,\ka )\  , \\
        \label{SJL_}
S_{J\, J-1}(\ka ,\kb ) &=& \sqrt{\frac{J}{2J+1}}
 \left( -1+ \frac{\ka +\kb}{J} \right) C_J(-\kb ,\ka ) \ .
\end{eqnarray}
In the case $J=0$ there is only one nonvanishing coefficient $S_{01}(\ka,\kb)=C_0(-\kb,\ka).$ The
coefficients $C_J(\kb ,\ka )$ are given by
\begin{eqnarray} \label{CJ}
C_J(\kb ,\ka ) = (-1)^{j_b+1/2} \sqrt{(2j_a+1)(2j_b+1)}
 \ThreeJ{j_a}{J}{j_b}{1/2}{0}{-1/2}
 \Pi (l_a,l_b,J) \, ,
\end{eqnarray}
where the symbol $\Pi (l_a,l_b,J)$ is unity if $l_a+l_b+J$ is even, and zero otherwise.

\section{Infrared divergent integrals}
\label{app:IR}

In this section we evaluate the infrared divergent integrals $J_{\alpha}$ with $\alpha = 2$ and
3, defined as
\begin{align}\label{eq:Ja}
J_{\alpha}(abcd) = \frac{i}{2\pi} \intinf d\omega\, \frac{\lbr ab|I_{\mu}(\omega)|cd\rbr}
  {(-\omega+i0)^{\alpha}}\,.
\end{align}
$I_{\mu}(\omega)$ is the electron-electron interaction operator with a finite photon mass $\mu$,
written in the Feynman gauge as
\begin{align}
I(\omega,x_{12}) = \alpha\,\big( 1 - \balpha_{1}\cdot\balpha_{2}\big)\,
\frac{e^{i\sqrt{\omega^2-\mu^2+i0}\,x_{12}}}{x_{12}}\,.
\end{align}

The integral over $\omega$ with $\alpha = 2$ is evaluated as
\begin{align}
\label{C2}
& \frac{i}{2\pi}\intinf d\omega \frac1{(-\omega+i0)^2}\frac{e^{i\sqrt{\omega^2-\mu^2+i0}\,x_{12}}}{x_{12}}
 = -\frac1{\pi x_{12}} \int_{\mu}^{\infty} \frac1{\omega^2} \sin \big(\sqrt{\omega^2-\mu^2+i0}\,x_{12}\big)
 =-\frac1{\pi x_{12}} \int_{0}^{\infty} dt \frac{t \sin tx_{12}}{(t^2+\mu^2)^{3/2}}
\nonumber \\ &
 =-\frac1{\pi} \int_{0}^{\infty} dt \frac{ \cos tx_{12}}{(t^2+\mu^2)^{1/2}}
 =-\frac1{\pi} \int_{0}^{\infty} dt \frac{ \cos tx_{12}-\cos t}{t} -\frac1{\pi} \int_{0}^{\infty} dt \frac{ \cos t}{(t^2+\mu^2)^{1/2}}
 = \frac1{\pi} \Big( \ln \frac{\mu}{2} + \gamma + \ln x_{12}\Big) + O(\mu)\,.
\end{align}

Therefore,
\begin{align}
J_{2}(abcd) =
  \frac{\alpha}{\pi} \Big( \ln \frac{\mu}{2}+\gamma\Big) \big[
 \lbr a|c\rbr\,\lbr b|d\rbr - \lbr a|\balpha|c\rbr \lbr b|\balpha|d\rbr\big]
  + \frac{\alpha}{\pi} \big< ab\big|\big( 1 - \balpha_{1}\cdot\balpha_{2}\big)\,\ln x_{12}\big|cd\big>\,,
\end{align}
where we dropped terms vanishing in the limit $\mu \to 0$. Analogously, we obtain
\begin{align}
J_{3}(abcd) =
  \frac{\alpha}{4\mu} \big[
 \lbr a|c\rbr\,\lbr b|d\rbr - \lbr a|\balpha|c\rbr \lbr b|\balpha|d\rbr\big]
  - \frac{\alpha}{4} \big< ab\big|\big( 1 - \balpha_{1}\cdot\balpha_{2}\big)\, x_{12}\big|cd\big>\,.
\end{align}

One can see that infrared divergences arise from terms of the type $J_{\alpha}(abab)$, since in
this case
$$
\lbr a|a\rbr\,\lbr b|b\rbr - \lbr a|\balpha|a\rbr \lbr b|\balpha|b\rbr = 1\,.
$$
We need also consider the case of $c = \widetilde{a}$ and $d = \widetilde{b}$, where the state
$\widetilde{n}$ has the same energy as $n$ but the opposite parity (e.g., $\widetilde{n} =
2p_{1/2}$ and $n = 2s$ for the point nucleus). Such states do not cause any infrared divergences
since $\lbr a|\widetilde{a}\rbr = 0$ due to orthogonality and the matrix element with $\balpha$
vanishes because of degeneracy in energy,
\begin{align}
\lbr a|\balpha|\widetilde{a}\rbr = \lbr a|i[{\cal H}_D, \bfr]|\widetilde{a}\rbr =
 i (\vare_a-\vare_{\widetilde{a}})\,\lbr a|\bfr|\widetilde{a}\rbr = 0\,.
\end{align}

\end{widetext}

\end{document}